\documentclass[prd,showpacs,nofootinbib,
12pt ]{revtex4}
\usepackage[cp1251]{inputenc}
\usepackage{amsmath}
\usepackage{amsfonts}
\usepackage{array}
\usepackage{amsthm}
\usepackage{latexsym}
\usepackage{epic}
\usepackage{graphicx}
\usepackage{eepic}
\usepackage[unicode=true]{hyperref}
\usepackage{psfrag}
\usepackage[mathcal]{euscript}

\usepackage{graphicx}

\newcommand{\ba}{\begin{array}}
\newcommand{\ea}{\end{array}}
\newcommand{\be}{\begin{equation}}
\newcommand{\ee}{\end{equation}}
\newcommand{\bea}{\begin{eqnarray}}
\newcommand{\eea}{\end{eqnarray}}
\newcommand{\e}{\mathrm{e}}
\newcommand{\nn}{\nonumber}
\newcommand{\lb}{\label}

\begin{document}

\title{``Triangular'' extremal dilatonic dyons}
\author{Dmitri Gal'tsov} \email{galtsov@phys.msu.ru}
\author{Mikhail Khramtsov} \email{hramcov.mihail@physics.msu.ru}
\affiliation{Department of Theoretical Physics, Moscow State
University, 119899, Moscow, Russia}
\author{Dmitri Orlov} \email{orlov_d@mail.ru }\affiliation{ITEP, Moscow, Russia}
\date{\today}

\begin{abstract}
Explicit dyonic dilaton black holes  of the four-dimensional
Einstein-Maxwell-dilaton theory are known only for two particular
values of the dilaton coupling constant $a =1,\sqrt{3}$, while for
other $a$ numerical evidence was presented earlier about existence
of extremal dyons in theories with the discrete sequence of dilaton
couplings $a=\sqrt{n(n+1)/2}$ with integer $n$. Apart from the lower
members $n=1,\,2$,  this family of theories does not have motivation
from supersymmetry or higher dimensions, and so far the above
quantization rule has not been derived analytically. We fill the
gap showing that this rule follows from analyticity of the dilaton
at the $AdS_2\times S^2$ event horizon, with $n$ being the leading
dilaton power in the series expansion. We also present its
generalization for asymptotically anti-de Sitter dyonic black holes
with spherical, planar and hyperbolic topology of the horizon.
\end{abstract}

\pacs{04.20.Jb, 04.50.+h, 04.65.+e}

\maketitle
 \section{Introduction}

Charged  black holes in four-dimensional Einstein-Maxwell-dilaton
(EMD) gravity  exhibit different features depending on the value of
the dilaton coupling constant $a$ entering the Maxwell term
$\e^{-2a\phi}F^2$ in the Lagrangian. Several particular values of
$a$ have higher-dimensional or supergravity origin. For $a=0$ the
dilaton decouples and the theory reduces to Einstein-Maxwell
(EM) system, which is the bosonic part of $N=2,\,D=4$ supergravity.
In this case the extremal dyons, defined geometrically as black
holes with the degenerate event horizon, saturate the supergravity
Bogomol'nyi-Prasad-Sommerfield (BPS) bound and have the $AdS_2\times
S^2$ horizon with zero Hawking temperature. For generic $a\neq 0$ only the
static purely electric or magnetic black holes are known analytically
\cite{Gibbons:1982ih,GiWi,Gibbons:1987ps,GHS}. In the non-extremal case they have two horizons,  the inner one being singular, so in
the extremality limit the event horizon becomes a null singularity
with vanishing Beckenstein-Hawking entropy and finite temperature
(small black holes). For $a=(p/(p+2))^{1/2}$ small black holes may
be interpreted as compactified regular non-dilatonic $p$-branes in the
$(4+p)$-dimensional theory \cite{Gibbons:1994vm}. The value $a=1$
corresponds to $N=4,\,D=4$ supergravity or dimensionally reduced
heterotic string effective action, in this case the dyonic solutions
are also known \cite{Gibbons:1982ih,Gibbons:1987ps} which are
non-singular in the extremal limit and  possess the $AdS_2\times S^2$
horizons. The last particular case, $a=\sqrt{3}$, corresponds to
dimensionally reduced $N=2,\,D=5$ supergravity; in this case the
static dyon solutions  also have the $AdS_2\times S^2$ horizon
structure in the extremality limit.

The {\em rotating} dyonic solutions are known analytically only for  $a=0$
and $a=\sqrt{3}$ \cite{Clement:1986bt,Rasheed:1995zv}. In the first
case it is the Kerr-Newman solution of  EM theory, while in the
second  these were derived  using the three-dimensional sigma-model
on the symmetric space $SL(3,R)/SO(2,1)$ corresponding to vacuum
five-dimensional gravity. The EMD theories with these two values of the
dilaton coupling exhaust the set of models reducing to
three-dimensional sigma-models on coset spaces
\cite{Galtsov:1995mb}, so from this reasoning there are no
indications on any particular  status of EMD theories with other
$a$. Meanwhile, as was shown numerically by Poletti, Twamley and
Wiltshire \cite{Poletti:1995yq}, the values $a=1,\,\sqrt{3}$ are
just the two lowest members $n=1,\, 2$ of the ``triangular''
sequence of dilaton couplings
 \be \lb{an}
a_n=\sqrt{n(n+1)/2}\,,
  \ee
selecting EMD theories in which numerical non-extremal dyonic
solutions exist with two horizons and admit the extremal limit. This
quantized sequence emerged as the result of numerical fitting of the
non-extremal solutions between the two horizons which looked
similar to solving a non-linear eigenvalue problem. Similar
picture persists in the Gauss-Bonnet gravity too \cite{Chen:2008hk}.
Meanwhile, no analytical justification of the rule (\ref{an}) was given so
far.

Some attempts are known to explore possibility of fake supergravity
embedding of the EMD theory with generic $a$. Using the Witten-Nester
construction, Gibbons et al. \cite{Gibbons:1993xt} were able to
derive the BPS-like inequality for arbitrary $a$. Meanwhile, as was
later shown by Nozawa and Shiromizu \cite{Nozawa:2010rf}, the
corresponding Killing spinor equations  do not imply the
bosonic equations as the integrability condition, unless
$a=0,\,\sqrt{3}$. So it looks unlikely that the  rule (\ref{an}) may
be interpreted as consequence of  some hidden supersymmetry.

Wiltshire and collaborators have also investigated dyonic solutions
with the cosmological constant
\cite{Wiltshire:1994de,Poletti:1994ww} and proved a no-go theorem for
asymptotically $dS_4$ black holes with $\Lambda>0$. On the contrary,
numerical evidence was presented about existence of asymptotically
$AdS_4$  black holes  for $\Lambda<0$ and certain values of $a$
which, however, were not found analytically.

In this paper  we investigate the nature of the dilaton coupling
quantization both for AF and asymptotically AdS dyons with various
topologies of the horizon. We show that the rule (\ref{an}) for AF
solutions follow from analyticity of the dilaton at the $AdS_2\times
S^2$ event horizon  and holds for all $n$ being the necessary
condition of existence of extremal solutions. We also derive the
generalization of the coupling quantization formula for
asymptotically $AdS_4$ dyons  with $AdS_2\times \Sigma_k$ horizons,
where $\Sigma_k=S^2, \,E_2$ or $H_2$. In all these cases the
solutions exist in the families of theories with $a$ depending on the
integer and the continuously varying $\Lambda$. The existence of
global solutions, both with  AF and AdS asymptotics, is confirmed
numerically, an analytic proof will be given elsewhere.

\section{ Setup} We choose the EMD lagrangian in the
form
 \be L=
R -2\Lambda - 2(\partial \phi)^2   - \e^{- 2a\phi}
  F^2\,,
 \label{action}
 \ee
and assume the static ansatz for the metric and the Maxwell one-form
:
 \begin{align}\label{ansatzM}
& ds^2 = -\e^{2 \delta}\, N\,  dt^2 +
\frac{dr^2}{N} +  R^2\,  d\Sigma^2_k\;,& \\
& A =-f\,dt-P\,\cos \theta\ d\phi\;,&
\end{align}
where $P$ is the magnetic charge. The metric on the
two-space $d\Sigma_{k}^2$ can be spherical, flat or hyperbolic:
\begin{equation}
d\Sigma_{k}^2 = \Bigg\{
 \begin{array}{c}   d\theta^2 + \sinh^2 \theta d\phi \,,\quad
  k=-1,\\  \!\!\!\!\!     d\theta^2  + \theta^2 d\phi
  \,,\quad\quad\quad
k=0,\\ \!\!\!\!\!   d\theta^2  + \sin^2 \theta d\phi \,,\quad
\;\;k=1,
\end{array}
\end{equation}
and the functions $f$, $\phi$, $N$, $R$ and $\delta$ depend on the
radial variable $r$  only. The equations of motions can be readily
found from the reduced one-dimensional lagrangian
 \be \label{L} \mathcal{L} = 2k \e^{\delta} + 2\e^{-\delta} \left (\e^{2\delta} N R\right)'
R' -2\ \phi'^2 \e^{\delta} N R^{2} - 2 \Lambda \e^{\delta} R^{2} +
2\ \e^{-2 a\phi} \left (f'^2 R^{2} \e^{-\delta} - \frac{P^2
\e^{\delta}}{R^{2}}\right)\,. \ee

Two particular gauges are
relevant: one is $\delta=0$ in which some exact solutions are known,
another is $R=r$, which is suitable for analytical derivation of the
quantization rules and for numerical calculations.
Choosing the gauge $R=r$ and solving Maxwell equations we find:
$$ f' =
 \frac{Q\;\e^{\delta+2 a\phi}}{r^{2}}\,,
 $$
 where the electric charge $Q$ is an integration
 constant. The remaining dilaton and Einstein
equations then read:
\begin{align}
 & \left(N \varphi' \e^{\delta} r^{2} \right)' = \frac{2a
 \e^\delta |PQ|}{ r^{2}}\sinh(2a\varphi)  \;,&\label{eqP}\\
 & \delta' =\varphi'^2 r\;,& \label{eqN}\\
 &\e^{ -\delta}\  \left (\!\e^{\delta} \!N r \right)'=
  k - \Lambda r^{2}  -
 \frac{2|PQ| \cosh(2a\varphi)}{ r^{2}} \,,& \label{eqD}
\end{align}
where the  shifted dilaton function is introduced
 \be\lb{fifi} \varphi=\phi-(\ln z)/2a\,, \qquad z=\left|\frac{P}{Q}\right|.
 \ee
The third Einstein equation is related to (\ref{eqN}-\ref{eqD})  via
the Bianchi identity. This system of equations possess a discrete
electric-magnetic duality \be \lb{sdu} P\leftrightarrow Q\,,\qquad
\varphi\leftrightarrow -\varphi \,,\ee
so we can restrict $a\geq 0$ without loss of generality.

For positive cosmological constant this system does not admit the desired
black hole solutions. In this case the
cosmological horizon exists outside the event horizon, while the
equations of motion imply zero value of some integral from a
non-negatively defined quantity over the space-like region between
two horizons \cite{Poletti:1995yq}. This argument holds only for
dyons with non-trivial dilaton, but fails for the singly charged
solutions \cite{HoHo} and for  the non-dilatonic dyons with equal
charges $P=Q$. In this latter case one has the well-known solution $\varphi \equiv 0,\;\delta \equiv
0$ and
 \be N = k - \frac{\Lambda}{3}r^2 - \frac{2M}{r} +
\frac{2Q^2 }{r^{2}} \,,\label{RNAdS} \ee where the mass $M$ enters as an integration constant. For $k=1$ this is the
Reissner-Nordstr\"om-de Sitter (anti-de Sitter) dyonic black hole
for   $\Lambda>0$ ($\Lambda<0$)   and the
pure Reissner-Nordstr\"om dyon for $\Lambda =0$. For $k=0 , -1$ and
$\Lambda<0$ the Eq. (\ref{RNAdS}) describes black holes with planar
and hyperbolic horizons.  The non-dilatonic solution with generic $Q$  and $M$ has
an  event horizon $r=r_h$ which is the highest root of the equation
$N(r_h)=0$. We are interested in {\em extremal} dyons with the horizons
$AdS_2\times\Sigma_k$, whose radius  $r_h$   (finite in the chosen
coordinates) satisfies the two equations:
 \be\lb{NN}
N(r_h)=0 \,,\quad N'(r_h)=0\,.
 \ee
Excluding the mass, we obtain an equation for $r_h$:
 \be\lb{r0}
 \left( k-\Lambda r_h^2\right)r_h^{2}=2Q^2\,,
 \ee
 while the second independent equation gives the mass in terms of
 $r_h$:
\be
 M=r_h \left(k-\frac{2\Lambda}{3}r_h^2\right) \,.
 \ee

If $Q^2 \neq P^2$, the dilaton is non-trivial, and then there is no
black hole solutions with positive cosmological constant
\cite{Poletti:1994ww}, so we are left with $\Lambda \leq 0$.

\section{Asymptotically flat dilatonic dyons} For $\Lambda =0$ one has only the
spherical geometry, $k=1$, and the asymptotic flatness (AF) implies
$N(\infty)=1\,,\, \delta({\infty})=0$ with the next leading terms
 \begin{align} & N  \sim  1 - \frac{2M}{r}+
\left(2|QP| \cosh(2a\varphi_\infty) + \Sigma^2\right)\frac{1}{r^{2}} , \label{AFN}\\
 &\e^\delta \sim 1-\frac{\Sigma^2}{2 r^{2}} ,\label{AFD}\\
 &\phi\sim \phi_\infty + \frac{\Sigma}{r} , \label{AFP}
 \end{align}
where $ M,\,\Sigma$ and $\phi_\infty$ ($\varphi_\infty$ is related to $\phi_\infty$
by (\ref{fifi})) are free parameters of the
local series solution. As expected, in global solutions the dilaton
charge $\Sigma$ is not an independent quantity: integrating the Eq.
(\ref{eqP}) one obtains the sum rule: \be\label{sigmaint}
 \Sigma =2a|QP|\int_{r_h}^{\infty}
 \frac{\e^\delta\sinh{2a\varphi}}{r^{2}}dr\;.
 \ee

It can be shown that for the AF solutions with the degenerate
horizon satisfying (\ref{NN}), there is a second constraint on the
charges, namely  the no-force condition. First, from the equations
of motion one can deduce that the quadratic form
 \be
I= \left( \frac12 N^2 \e^{2\delta} \varphi'^2  + \frac{1}{8}
\e^{-2\delta}(N\ \e^{2\delta})'^2 \right)r^{4}-
 |QP|N \e^{2\delta}
\cosh(2a\varphi)
 \ee
is conserved on shell:
 \be
\frac{dI}{dr}=0\,,
 \ee
(similar expression in the gauge $\delta =0$ was given in
 \cite{Poletti:1995yq}).
Substituting $r=r_h$  one finds that in view of (\ref{NN})
this integral actually has zero value, $I=0$.
Then substituting the asymptotic expansions (\ref{AFN}-\ref{AFP}) we
obtain from $I=0$ the no-force condition:
 \be
 M^2+\Sigma^2 =Q_\infty^2+P_\infty^2\,, \lb{no-force}
 \ee
where $Q_\infty=Q\e^{2a\phi_\infty}, P_\infty=P\e^{-2a\phi_\infty}.$
For known exact solutions $\phi_\infty=0$, so the asymptotic charges
coincide with the initial ones. Constructing the numerical solutions,
we will impose the same condition adjusting the horizon data.

An exact extremal dyon solution  with $a=1\, (n=1)$ in this gauge reads
 \be
\e^{-2\delta}=1+\frac{\Sigma^2}{r^2}\,,\quad N=
\left(1-\frac{2M}{r^2}\sqrt{r^2+\Sigma^2}+\frac{Q^2+P^2+\Sigma^2}{r^2}\right)\,,\quad
 \e^{2\varphi}=\left|\frac{Q}{P}\right|\cdot
 \frac{\sqrt{r^2+\Sigma^2}+\Sigma}{\sqrt{r^2+\Sigma^2}-\Sigma}\,.
  \ee
It has a horizon at $r_h=\sqrt{M^2-\Sigma^2}$ and the dilaton charge satisfies
 \be \lb{Si1} 2M\Sigma=P^2-Q^2\,.
  \ee
Another exact dyon solution is known for $n=2,\, (a=\sqrt{3})$
\cite{Gibbons:1982ih,Gibbons:1987ps}. The corresponding dilaton
charge satisfies the following formula
 \be
 \frac{Q^2}{\Sigma-a M}+\frac{P^2}{\Sigma+a
 M}=\frac{1+a^2}{2a^2}\Sigma\,, \label{consan}
 \ee
which is valid both for $a=a_1\,,a_2$.

Further information about higher $n$ dyons can be extracted from
known exact solutions for {\em singly charged} black holes. This provides
us with some knowledge about the limiting point $z=0$ (an opposite
limit $z=\infty$ can be explored via the discrete electric-magnetic
duality (\ref{sdu})). These solutions look simpler in the gauge
$\delta=0$, the corresponding equations of motion being
 \bea
&&\left(N \varphi' R^{2} \right)' = \frac{2a
 |PQ|}{ R^{2}}\sinh(2a\varphi)  \;, \\
&& R'' +\varphi'^2 R=0\;,\\
&&  \left (N R \right)'=
  k - \Lambda R^{2}  -
 \frac{2|PQ| \cosh(2a\varphi)}{ R^{2}} \,.
\eea The electrically  charged solution  valid for all $a$ in the
general non-extremal case reads:
 \begin{align}\lb{sta}
R^2=\rho^2f_-^{1-\gamma}\,,\quad N=f_+f_-^{\gamma}\,,\quad
\e^{2a\varphi}=f_-^{\frac{2a}{1+a^2}}\,,\quad
\gamma=\frac{1-a^2}{1+a^2}
 \,,
 \end{align}
with $ f_\pm=1-r_\pm/\rho$, where we denoted the radial coordinate
as $\rho$ to distinguish it from $r$ in the gauge used in the Eqs.
(\ref{eqP}-\ref{eqD}). The mass, the electric charge and the dilaton
charge are related to $r_\pm$ via
 \be\lb{char0}
r_+ r_-=\frac{2Q^2}{1+\gamma},\,\quad 2M=r_++\gamma r_-\,,\quad
\Sigma = -\frac{a}{1+a^2} r_-\,.
 \ee

Note that our system of equations has special points of two kinds:
the zeroes of the function $R(\rho)$, which correspond to the
curvature singularity, and the zeroes of $N(\rho)$ for which $R\neq
0$, where the space-time geometry is regular. Generic behavior of the metric
functions and the dilaton near the curvature singularity is
non-analytic (unless $a=1$), while in the second case it {\em is} analytic. This is
clearly seen from the singly charged solution (\ref{sta}) which has two zeroes of
$N(\rho)$ at $\rho=r_{\pm}$, with $r_-$ being a zero of $R(\rho)$
too. The dilaton is analytic at the regular horizon
 $\rho=r_+$, but non-analytic in the singularity
$\rho=r_-$. This can be expected for dyonic solutions as
well.
In the extremal limit $r_+=r_-$ the dilaton is therefore singular at
the horizon, but this does not influence its asymptotic behavior. So
the dilaton charge  is still finite and given by (\ref{char0}). In
the extremal limit it reads:
 \be \lb{SiQ}\Sigma = -\frac{a\,Q}{\sqrt{1+a^2}} \,.
 \ee This expression is expected to match the
corresponding limit of dyonic solutions.

Now let us look for extremal dyon solutions with the regular horizon
for arbitrary $a$. It turns out that already {\em the
local} power series solution in the vicinity of the degenerate
horizon implies a constraint on $a$.
Coming back to the gauge $R=r$, in which  curvature singularity
is at $r=0$, and assuming (\ref{NN}) we will have in the leading order:
 \bea
&& N = \nu  x^2  + O(x^3), \qquad x=(r-r_h)/r_h \,,\label{expNeAF}\\
&& \varphi = \varphi_h + \mu   x^n + O(x^{n+1})\,, \label{expPeAF}
 \eea
where $\mu$ and $\nu$ are dimensionless parameters, and $n$ is an
integer. Substituting this into the Eq.~(\ref{eqN}) we find
 \be
\delta = \delta_h + \frac{\mu^2 n^2}{(2n-1)} x^{2n-1} + O(x^{2n})
\,,
 \ee and therefore $\e^\delta$ is finite and continuous at the
horizon. So the leading term of the l.h.s. of the Eq.~(\ref{eqP}) is
  \be \left (N \varphi' \e^\delta r^{2} \right )' =
 \nu \mu n (n+1)\e^{\delta_h} r_h^{2}x^n  + O(x^{n+1}) \label{+6}\;.
  \ee
 This is zero at $x=0$ for any $n$,
so looking at the r. h. s. of Eq.~(\ref{eqP}) we immediately find that one must have
$$\varphi_h=0\,.$$
Then the linear in $x$ term
at the r. h. s. of (\ref{eqP}) agrees with that at the l.h.s. and we
find:
 \be \nu n(n+1) =  {4 a^2 |QP|}/{r_h^{2}}.\lb{+8}
 \ee
 Now
consider the equation (\ref{eqD}) in the vicinity of  $r \sim r_h$.
One sees that the l.h.s. is linear in $x$, so the r.h.s. has to be
linear in $x$ too. Vanishing of the constant terms for $\Lambda=0,
k=1$  gives an equation for $r_h$:
 \be r_h^2=2|QP|\,, \label{+10}
  \ee
while expanding $ r^{-2} = r_h^{-2} (1-2x ) +...$ and equating the
linear in $x$ terms we obtain $\nu =1 $. Substituting this and
(\ref{+10}) into (\ref{+8}), we arrive at
 \be a^2 = a^2_n=\frac{n(n+1)}{2}
\,.\label{aquant}
 \ee This is the necessary condition for existence
of the AF regular extremal dilatonic dyons which coincides with
(\ref{an}). It shows, in particular, that such solutions do not
exist for $a<1$, except for $a=0$.

The main characteristic of new solutions is the dependence of the dilaton charge $\Sigma$ on $z$. This can be extracted  from the numerical solutions which must
fit the  no-force condition (\ref{no-force}).
To construct numerical solutions of the system
(\ref{eqP})-(\ref{eqD}) on the interval $r_h< r<\infty$ we have to
set initial conditions at some tiny step away from the
horizon $r_h$   which is the singular point
of the system of differential equations. To perform this with a given accuracy
we need to keep more terms in the series expansions:
 \bea && N
= \sum_{k=2}^{2n}(-1)^{k} (k-1)\ x^{k}-n (\mu^2 +2)\ x^{2n+1} +
O(x^{2n+2})\,,\nn \\\nn
&& \varphi = \mu\  x^n -n \mu\  x^{n+1} + \frac{n(n+1)}{2}\  x^{n+2} + O(x^{2n})\,,\label{horAF} \\
&& \delta = \delta_h + \frac{\mu^2 n^2}{(2n-1)}\   x^{2n-1} +
O(x^{2n}) \,,  \nn
 \eea
This local solution has two free parameters: $\mu$ and $\delta_h$, from which the latter is trivial
because of the symmetry of the Eqs.~(\ref{eqP}-\ref{eqD}) under translations of $\delta$.
At infinity we have  asymptotic expansions in terms of the
physical mass $M$, the dilaton charge and the electric and magnetic
charges (\ref{AFN})-(\ref{AFP}). Actually, the mass can be extracted
from the no-force condition (\ref{no-force}), and one of the charges
may be set equal to one defining the length (mass) scale. So the
only remaining parameter which has to be computed from numerical
solutions is the dilaton charge as a function of the ratio $z=|P/Q|$. In view of the electric-magnetic
duality it will be enough to consider the interval $z\in[0,\,1]$, i.e. the electrically dominated solutions. For them the dilaton charge $\Sigma$ is negative.

\begin{figure}
\begin{center}
\includegraphics[width=370pt]{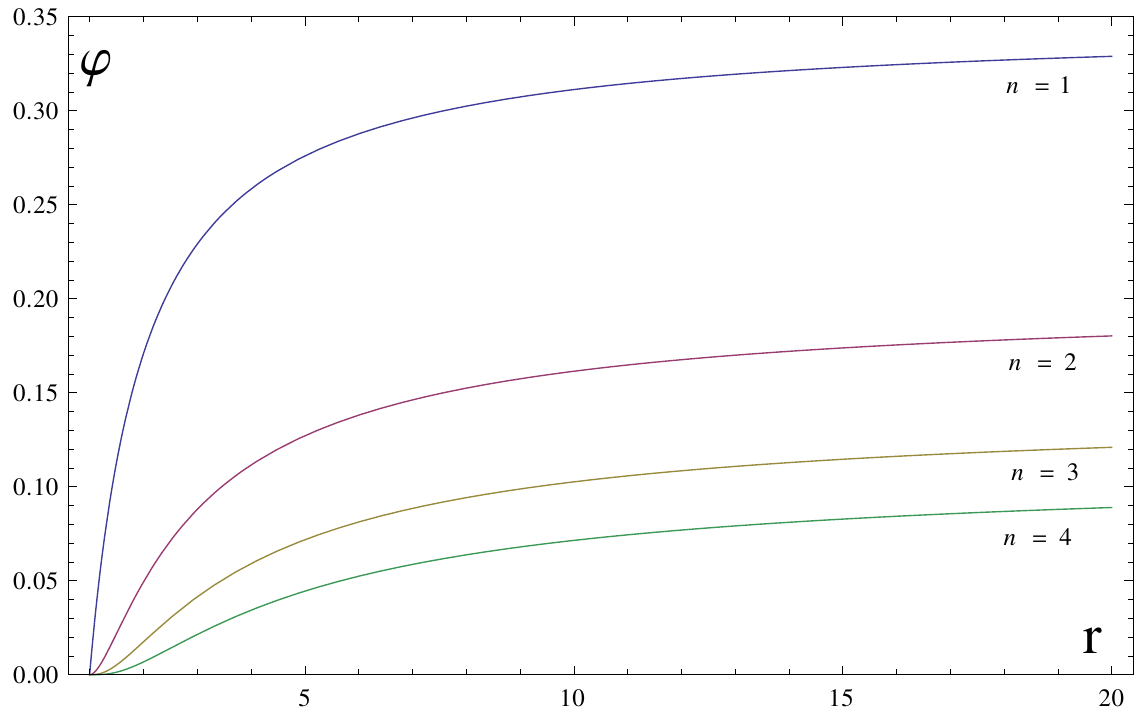}
 \caption{Behavior of the dilaton function $\varphi(r)$ of   asymptotically flat electrically dominated dyons for  $n=1, 2, 3, 4$. The
curves start from zero and approach the values given by  Eq.~(\ref{varphias}).} \label{solution}
\end{center}
\end{figure}

\begin{figure}
\begin{center}
\includegraphics[width=370pt]{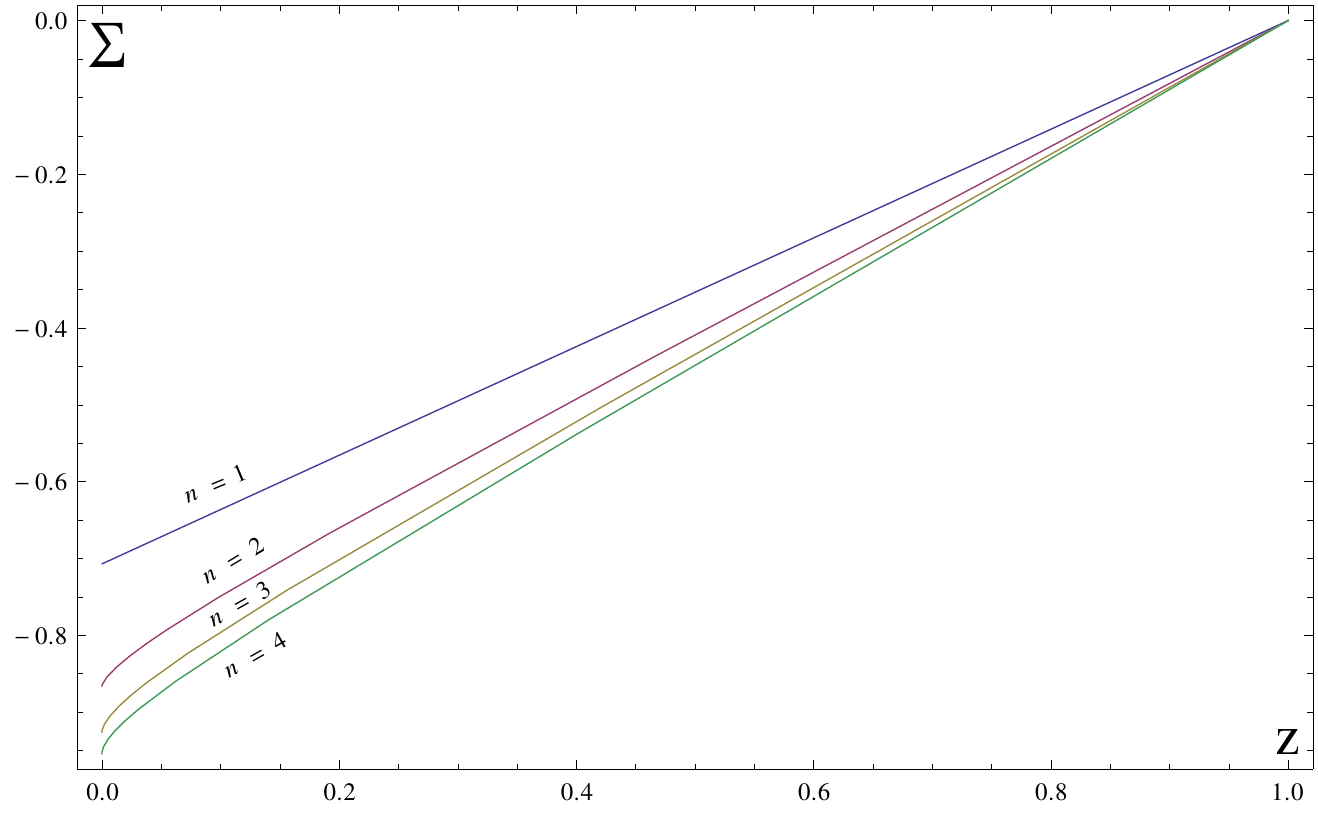}
 \caption{Dilaton charges $\Sigma$ defined by   Eq.~(\ref{sigmaint}) for electrically dominated AF dyons as functions of $z=|P/Q|$ for
$n=1, 2, 3, 4$. The
curves start with the values given by   Eq.~(\ref{SiQ}) at $z=0$ and tend
to zero at $z=1$, corresponding to non-dilatonic solutions.} \label{sigmaAF1}
\end{center}
\end{figure}

For each given $z$ there is a set of $\mu$ yielding asymptotically flat solutions with different
values of the dilaton at infinity,
from which we extract one corresponding to $\phi_{\infty}=0$, or in terms of $\varphi$:
\be\lb{varphias}
\varphi_\infty=\frac1{2a}\ln\left|\frac{Q}{P}\right|. 
\ee
In this case the physical charge $Q_{\infty}$
coincides with the initial $Q$, so the no-force condition (\ref{no-force})  reduces to $M^2+\Sigma^2=Q^2+P^2$.
A series of lower-$n$ numerical solutions for the dilaton $\varphi(r)$ are shown on   Fig.~\ref{solution}.
The metric functions $N(r)$ and $\delta(r)$ are monotonous, increasing  from zero values at the horizon to unity  at infinity, so we do not show them here. The solutions with various $z$ are then used to calculate  $\Sigma(z)$ via the sum rule
(\ref{sigmaint}).
The resulting curves for lower $n$ are shown on
 Fig.~\ref{sigmaAF1}. They interpolate between the $z=0$ values $\Sigma_n(0)$ given by (\ref{SiQ}) with $a$ equal to (\ref{an})
and zero value at $z=1$. Note that the sequence $\Sigma_n(0)$ converges to $-Q$ as $n\to\infty$.

\section{Asymptotically AdS dyons}
For $\Lambda <0$ we keep the
topological parameter $k=1,0,-1$ arbitrary. Then the  asymptotic
behavior in the leading order is given by:
 \begin{align} & N \sim  \frac{ r^2}{l^2}  +  k   -  \frac{2 M}{r}  +
\frac{Q_\infty^2+P_\infty^2}{r^2} \label{LN}
,\\
&\phi  \sim \phi_\infty + \frac{\hat\Sigma}{r^{3}} ,\label{LP}\\
&e^{\delta } \sim 1-\frac{3\hat{\Sigma}^2}{2 r^{6}} \label{LD} \,,
\end{align}
where the AdS curvature radius is given by
 \be\lb{l}
 l^2=-\frac{3}{\Lambda }\,.
 \ee
The modified dilaton charge $\hat{\Sigma}$ has the dimension of
(length$^3$) and satisfies the sum rule
 \be\label{sigmaintla}
 \hat{\Sigma} = \frac{2 a|QP|}{\Lambda}\int_{r_h}^{\infty}
 \frac{\e^\delta\sinh{2 a\varphi}}{r^{2}}dr\;.
 \ee

To generalize the coupling quantization condition (\ref{aquant}) we
follow the same strategy keeping track for $\Lambda$ and $k$
terms the Eq. (\ref{eqD}). Then we obtain the following equation:
\be \lb{r_hL}
  r_h^2 (k-\Lambda r_h^2)=2|PQ| \,,
  \ee
  leading to
  \be
r_h^2 = -\frac{1}{2\Lambda} \left(-k\pm\sqrt{k^2-8\Lambda |PQ|}\right)\,, \lb{r_hAdS}
 \ee
where we have to choose the upper
sign. Note that for $k=-1$ the horizon radius remains finite  if one of the charges
or both of them  are zero:
 \be \lb{rh0}
 r_h^2=l^2/3\,.
 \ee
This case may be therefore attributed
either to vacuum solution with the degenerate horizon, previously mentioned in \cite{Charm},
or to a singly charged planar black hole.

  Equating the
linear in $x$ terms in  (\ref{eqD}) we get
 \be \nu =k-2 \Lambda
r_h^2\,.
 \ee
 Substituting this to the Eq.(\ref{+8}), we obtain
 \be
 a^2 =\frac{n(n+1)}{2} \left(1- \frac{\Lambda r_h^2}{k-\Lambda
r_h^2}\right) \lb{aquantL}\,,
 \ee
where $r_h$ is the solution of
(\ref{r_hL}). The novel feature of this condition is the dependence
of the critical $a$ on the ratio of the horizon radius to the AdS curvature
radius $l$: $|\Lambda |r_h^2=3r_h^2/l^2$, except for the planar case $k=0$.
Actually the dilaton constant is a {\em parameter of the theory} within which
we are looking for dyonic solutions. So this relation should be regarded as determining
the horizon radius for given $\lambda$ and $a$. It has
multiple  branches labeled by $n$. Substituting (\ref{r_hAdS}) into  (\ref{aquantL})
we find another useful formula
 \be \lb{alk}
 a^2 = n(n+1) \left(1+ \frac{k}{\sqrt{k^2 + 24|QP|/l^2} }\right)^{-1} \,.
 \ee
In the limit $l^2\to \infty$ we come back to the triangular number (\ref{an}).
In the planar case $k=0$ we obtain
\be
a^2 =  n(n+1)\,, \label{k=0a}
\ee
which is also the limit $|QP|/l^2\to \infty$ for the spherical case $k=1$.
For $n \geq 2$ the branches for $k=1$  with different $n$  start to overlap thus allowing for spherically symmetric extremal dyons for any $a \geq a_2$.
 In the hyperbolic case $k=-1$ the Eq.~(\ref{aquantL})  exhibits divergence at  $r_h^2$ given by (\ref{rh0}), corresponding to the uncharged (or singly charged) solution.

The modified dilaton charge $\hat{\Sigma}$ can be found numerically using the same strategy as in the AF
case.  The initial data are obtained from the   near-horizon expansions:
\bea
&& N(r) = -\frac{a^2 r_h^2\ \Lambda}{a^2 - a^2_n} x^2 + \sum_{m=3}^{2n} (-1)^{m+1} \frac13 \frac{\Lambda r_h^2(2a^2 + n(n+1)(\frac32 m -1)}{a^2-a_n^2}x^m +
 O(x^{2n+1})\,,\nn \\
&& \varphi(r) = \mu x^n + \sum_{l=n+1}^{2n} \frac{\mu\ \Pi_l (a)}{a^l} x^l + O(x^{2n})\,,\\
&& \delta = \delta_h+ \frac{2\mu^2 n^2 }{2n-1} x^{2n-1} + O(x^{2n}) \,,
\eea
where $\Pi_l(a)$ are polynomials of the degree $l$.
Boundary conditions at  infinity are specified by the expansions (\ref{LN})-(\ref{LD}).
Contrary to the AF case, the system no longer
has the scaling symmetry, which makes numerical procedure more tedious.

\begin{figure}
\begin{center}
\includegraphics[width=370pt]{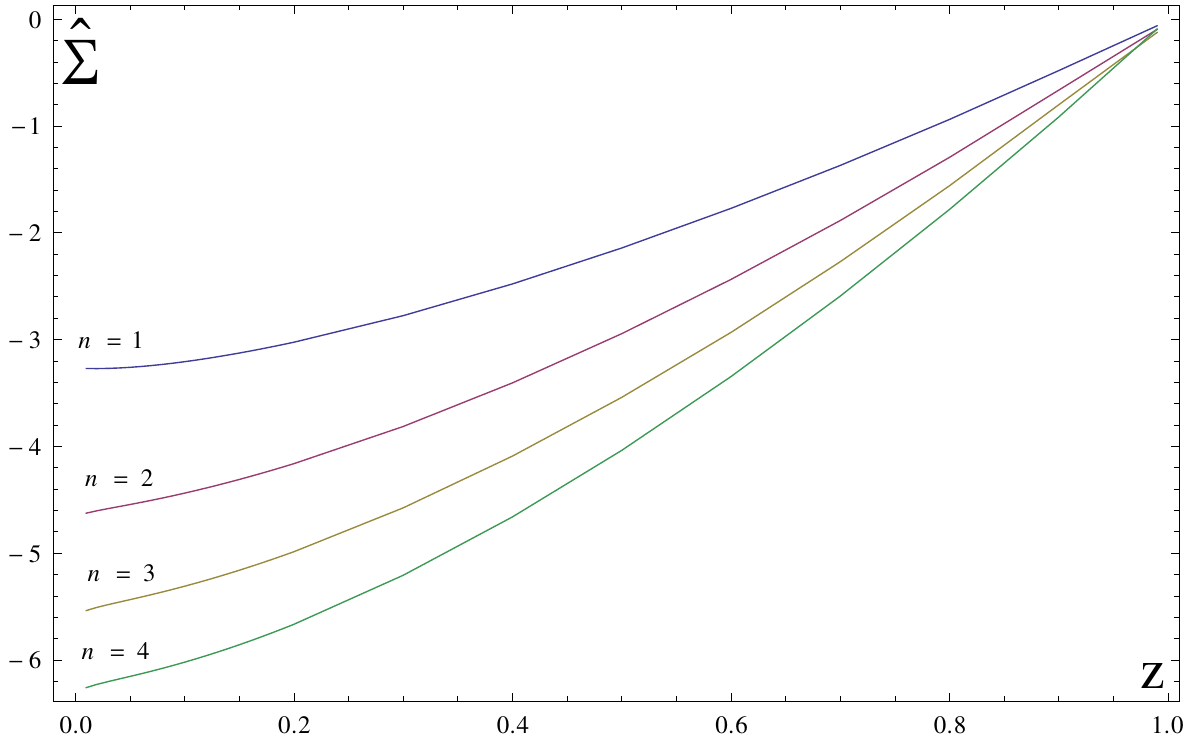}
\caption{Modified dilaton charges $\hat{\Sigma}$
(\ref{sigmaintla})   for electrically dominated asymptotically AdS dyons with  spherical horizons, $k=1$.}
\label{sigmaL}
\end{center}
\end{figure}

\begin{figure}
\begin{center}
\includegraphics[width=370pt]{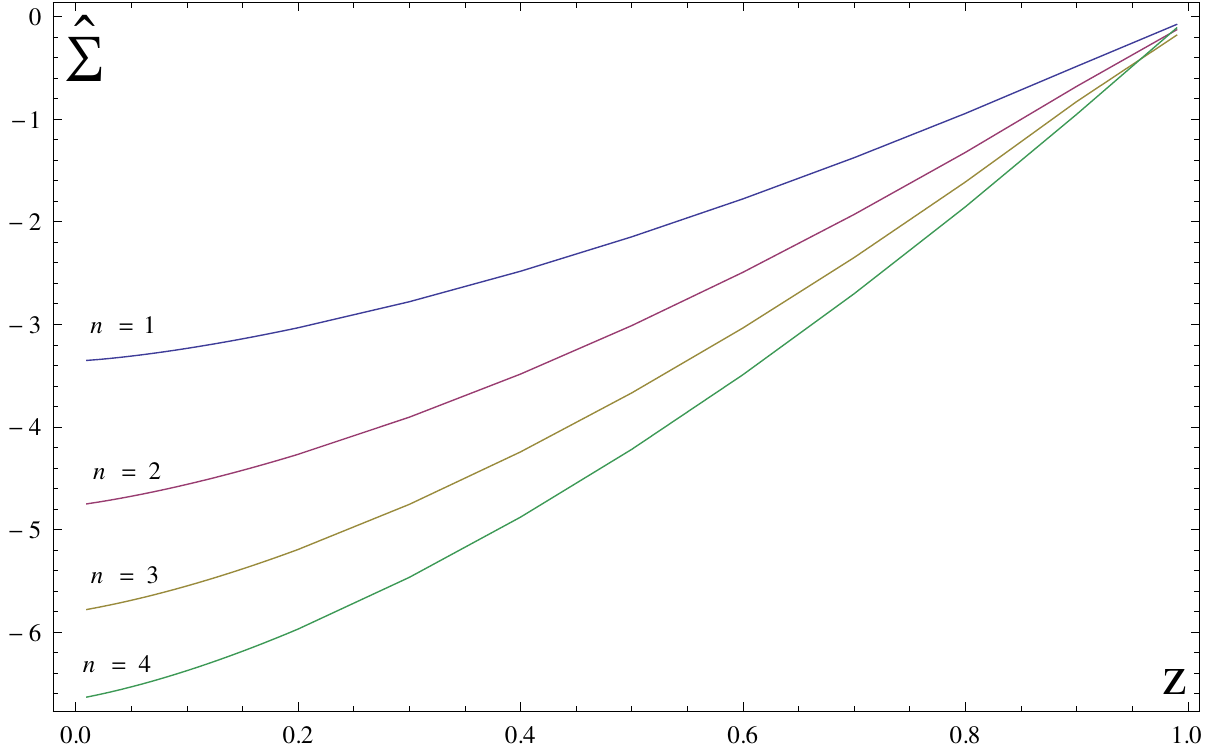}
\caption{Modified dilaton charges $\hat{\Sigma}$
(\ref{sigmaintla})  for electrically dominated asymptotically AdS dyons with planar horizons, $k=0$.}
\label{sigmaL0}
\end{center}
\end{figure}

\begin{figure}
\begin{center}
\includegraphics[width=370pt]{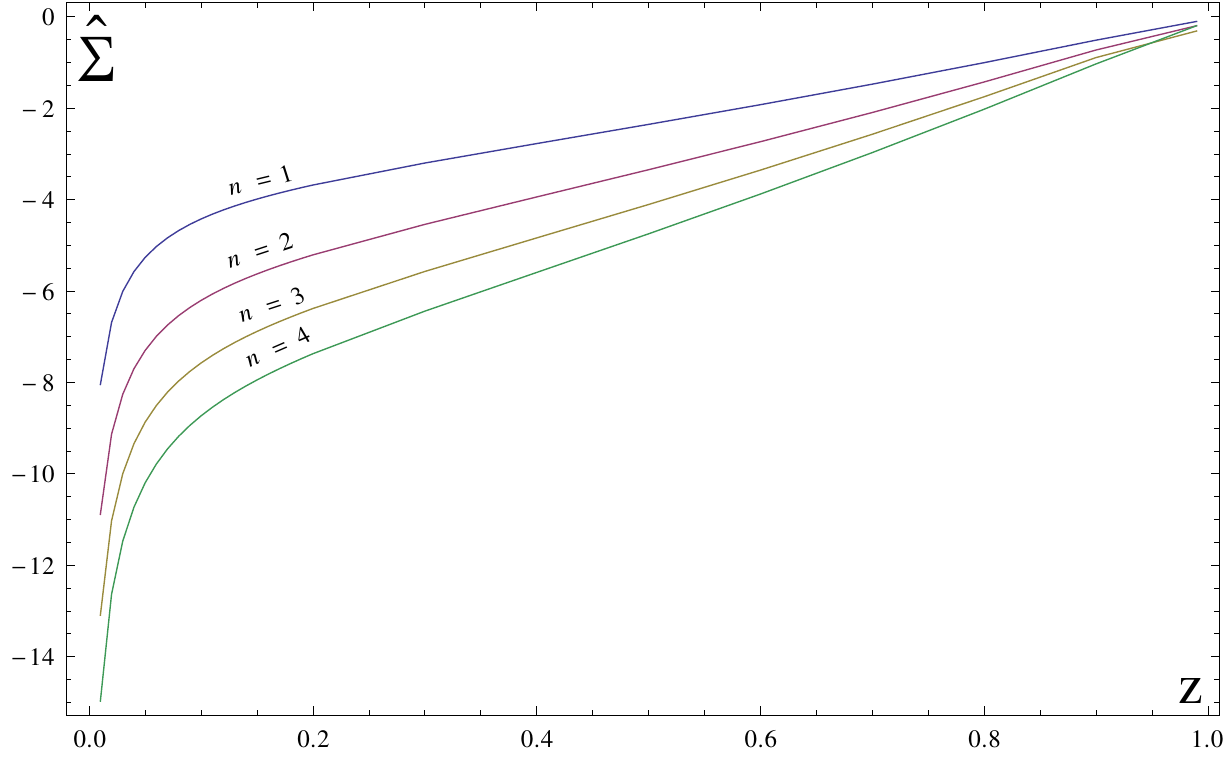}
\caption{Modified dilaton charges $\hat{\Sigma}$
(\ref{sigmaintla})   for electrically dominated asymptotically AdS dyons with  hyperbolic horizon, $k=-1$.  }
\label{sigmaL-1}
\end{center}
\end{figure}

The results are shown on Figs.~(\ref{sigmaL}-\ref{sigmaL-1}).
Note that for the hyperbolic horizon
 the dilaton charge $\hat{\Sigma}$ tends to $-\infty$ at $z=0$. This follows
directly from the formula (\ref{aquantL}). Since the radius of the
horizon tends to $-1/\Lambda$, the denominator in brackets explodes,
which leads to the divergence of $\hat{\Sigma}$  through the sum rule
(\ref{sigmaintla}).

\section{Conclusions} Our main result is an analytic derivation of
the ``triangular'' quantization rules for the dilaton coupling $a$
in EMD($\Lambda$) theory as necessary condition of existence of
regular extremal dilatonic dyons. These rules follow from the purely
local analysis of analyticity of the dilaton at the  $AdS_2\times
S^2$ event horizon. The integer $n$ is shown to be
the leading power index in the Taylor expansion of the dilaton function.
In the AF ($\Lambda=0$) case our result reproduces the discrete
sequence found by Poletti, Twamley and Wiltshire
\cite{Poletti:1995yq} in a slightly different setting. In the
asymptotically AdS case our formulas are entirely new and applicable
to topological solutions as well.  Dyons with hyperbolic topology
are shown to make contact with the uncharged solutions with AdS
asymptotic \cite{Charm}.

We expect possible generalization of our results to dyonic branes
\cite{Clement:2005vn}. A challenging task is to find analytical
formulas for triangular dyons, some evidence in favor of such a
possibility was presented in \cite{Chen:2008hk}.

 The authors thank Gerard Cl\'ement for
stimulating discussions  and reading the manuscript.  This work was
supported   by the RFBR grant  14-02-01092-a.

\newpage


\end{document}